\documentclass[letterpaper]{JHEP}
\usepackage{epsfig}
\usepackage{graphicx}
\usepackage{amssymb,amsmath}
\usepackage{epstopdf}

\def\be{\begin{eqnarray}}
\def\ee{\end{eqnarray}}

\title{When D-branes Break}

\author{Oren Bergman\\
Department of Physics\\
Technion, Israel Institute of Technology\\
Haifa 32000, Israel\\
\email{bergman@physics.technion.ac.il}}

\author{Gilad Lifschytz\\
Department of Mathematics and Physics and CCMSC \\
University of Haifa at Oranim\\
Tivon 36006, Israel \\
\email{giladl@research.haifa.ac.il}}

\date{}

\abstract{We analyze the possible configurations of D-branes breaking on other 
D-branes. We describe these configurations in the context of a brane-antibrane 
effective theory in two ways. 
First as a tachyon configuration representing a non-trivial bundle over the sphere 
surrounding the end of the brane a la Polchinski,
and second in terms of tachyon solitons using homotopy theory.
Surprisingly, in some cases there are topologically stable configurations
of broken branes.}

\begin{document}

\section{Introduction}

In string theory there are several circumstances in which branes can terminate on other branes.
Strings can of course end on D-branes, and using S and T-duality one can show
that a D$p$-brane can end on a D$(p+2)$-brane, and that a D$p$-brane with $p\leq 5$
can end on an NS5-brane. These configurations were shown to be consistent with
charge conservation by Strominger \cite{Strominger:1995ac}.
For a $p$-brane to end on a $q$-brane requires a coupling between the $(p+1)$-form
gauge field and the worldvolume gauge field on the $q$-brane.
This in turn implies that the end of the $p$-brane is charged under the worldvolume gauge field
on the $q$-brane.
For example, a string can end on a D$p$-brane because there is a term
$*F\wedge B$ in the expansion of the DBI action on the D$p$-brane.
This coupling implies that the end of the string is charged electrically
under the worldvolume gauge field.
Similarly, a D$p$-brane can end on a D$(p+2)$-brane because there is a term
$F\wedge C_{p+1}$ in CS action on the D$(p+2)$-brane.
This gives the $(p-1)$-dimensional end of the $p$-brane a magnetic charge under
the worldvolume gauge field in the D$(p+2)$-brane.
The full CS action on a D$q$-brane contains more terms coupling the worldvolume
gauge field to all lower rank RR fields,
\be 
\label{CS_action}
S_{CS} =  C_{q+1} + C_{q-1}\wedge F 
+ {1\over 2}C_{q-3}\wedge F\wedge F + \cdots \,,
\ee
which suggests that any D$p$-brane can end on a D$q$-brane with $p<q$,
as long as the end of the $p$-brane carries a charge of the form
\be
\label{end_charge}
 \int_{S^{q-p}} \mbox{Tr}\,F\wedge F\wedge\cdots\wedge F\,.
\ee
These possibilities were mentioned briefly in \cite{Copeland:2003bj}.
However this requires a large enough gauge group such that the
gauge bundle with the above Chern class is non-trivial,
so we need several coincident $q$-branes for the $p$-brane to end.

Charge conservation is a necessary condition for these configurations to exist,
but it isn't obviously sufficient. 
Even if they do exist, they may not be stable.\footnote{Existence and stability are separate questions.
For example, neutral non-BPS D$p$-branes of the "wrong" dimension $p$ {\em exist} in Type II
string theory ($p$ odd in Type IIA and even in Type IIB), but they are not {\em stable}.
On the other hand neutral D$p$-branes with the same $p$ as the charged BPS D-branes
{\em do not exist}.}
Polchinski has recently provided an explicit construction for the case of a 1-brane
ending on 9-branes in Type I string theory, in terms of an effective tachyon field
theory on a system of 9-branes and anti-9-branes \cite{Polchinski:2005bg}.\footnote{In 
\cite{Hashimoto:2001rj} Hirano
and Hashimoto constructed a D8-brane with a D6-brane ending on it
as a non-uniform 8-brane soliton in unstable D9-branes in Type IIA string theory.}
The basic ingredient is Sen's construction, and Witten's generalization, 
of D$p$-branes as topological solitons in a higher-dimensional 
brane-antibrane system \cite{Sen:1998tt,Witten:1998cd}.
In particular the 1-brane in Type I string theory corresponds to a co-dimension
eight tachyon configuration in the effective field theory of eight 9-brane-anti-9-brane
pairs, which asmptotically behaves as $T(x)\sim \Gamma\cdot x$.
Adding eight more 9-branes, Polchinski showed that the new
system admits a tachyon configuration that describes a semi-infinite 1-brane, namely a configuration 
that reduces to the above behavior on one side of the coordinate axis, and to zero on the other 
side.\footnote{Other aspects of asymmetric brane-antibrane systems were studied
in \cite{Hashimoto:2005qh,Ishida:2006tj}.}
The obvious interpretation is that the 1-brane ends on the eight 9-branes that remain after
the tachyon condenses. This agrees with the requirement (\ref{end_charge}), since $O(8)$ 
admits a non-trivial bundle over $S^8$.

We will generalize this construction to all the D$p$-branes of Type I
and Type IIB string theory, and show that all the (stable) $p$-branes
can end on 9-branes, provided there are enough 9-branes.
This will be in accord with the charge conservation condition (\ref{end_charge}).
We will also show that the existence of a configuration where a $p$-brane
ends on 9-branes is connected to the existence of a $(p-1)$-dimensional
soliton in a Higgs phase of the 9-brane-anti-9-brane model, where the extra
9-branes are separated (by Wilson lines) from the brane-anti-brane pairs.
This soliton sits at the boundary of the $p$-brane in the 9-brane, and carries the
charge of (\ref{end_charge}). 
%The two descriptions of the open $p$-brane,
%with and without the Higgs, agree on the conditions for branes to end,
%but they are valid in different regimes. We conjecture that they are connected
%by a phase transition.
Interestingly we will find examples of $p$-branes with both ends on 9-branes
which are topologically stable.

In section 2 we will review Polchinski's construction, and generalize it to
$p$-branes ending on 9-branes in Type I and Type IIB string theory.
In section 3 we will analyze the Higgs phase of the brane-anti-brane model,
and describe the end of the $p$-brane as a $(p-1)$-dimensional topological soliton.
Section 4 contains our discussion.

%%%%%%%%%%%%%%%%%%%%%%%%%%%%%%%%%%%%
%%%%%%%%%%%%%%%%%%%%%%%%%%%%%%%%%%%%

\section{Tachyon model}
%\section{Open D-branes as sphere bundles}

\subsection{Open Type I D1-brane}
In \cite{Polchinski:2005bg} Polchinski constructed a tachyon configuration 
which describes a semi-infinite D1-brane in Type I string theory.
He considered a sub-system of 16 9-branes and 8 anti-9-branes, with
a gauge group $O(16)\times O(8)$.
Since the tachyon is a bi-fundamental field it can be thought of as a map
from $R^8$ to $R^{16}$. The vev of the tachyon, up to a multiplicitive constant,
is given by
\be
 T_0 = \left[
 \begin{array}{cc}
 I_{8\times 8} \\
 0_{8\times 8}
\end{array}
\right] \; \mbox{mod}\; O(16) \,,
\ee
so it reduces to a map between the spheres
\be
T_0: S^7 \rightarrow S^{15} \,.
\ee
Polchinski then considers a tachyon configuration which asymptotically on
an $S^8$ surrounding the origin is given by the third Hopf map
\be 
\label{bundle4}
T(\Omega): S^7\rightarrow S^{15}\rightarrow S^8 \,.
\ee
%A semi-infinite D1-brane corresponds to a configuration in which the Tachyon
%approaches a vacuum value everywhere on an infinite $S^8$ surrounding the 
%endpoint, except at one point where it is singular. The asymptotic form of
%the tachyon $T(\Omega)$ therefore corresponds to the sphere bundle
The configuration is explicitly
\be
\label{open_D1brane}
T(\Omega) = {r+ \sum_{i=1}^9\gamma^i x^i\over\sqrt{2r(r+x^9)}} T_0 \,,
\ee
where $\gamma^i$ are the $SO(9)$ gamma matrices.
%Such a bundle exists, and is given by
%the fourth Hopf bundle, which is defined by the Hopf map
%\be 
%x^i=  y^T \gamma^i y \,,
%\ee
%where $x\in R^9$, $y\in R^{16}$ and $\gamma^i$ are the nine $SO(9)$ Gamma matrices.
%By rewriting it slightly as 
%\be
%T(\Omega) = {1\over\sqrt{2r}}\left[
%\begin{array}{cc}
%(r+x^9)^{1/2} I \\
%(r+x^9)^{-1/2} \sum_{i=1}^8 \Gamma^i x^i 
%\end{array}\right]
%\ee
Near the positive $x^9$ axis
\be
T\sim \left[
\begin{array}{cc}
I \\
0
\end{array}
\right] = T_0 \,,
\ee
which is the vacuum, and near the negative 
$x^9$ axis
\be
T\sim \left[
\begin{array}{cc}
0 \\
\sum_{i=1}^8 \Gamma^i x^i/|x|
\end{array}
\right]\,,
\ee
where $\Gamma^i$ are the $SO(8)$ gamma matrices.
This is precisely the asymptotic form of the tachyon for the D1-brane.
The configuration (\ref{open_D1brane}) therefore describes a semi-infinite D1-brane,
with an endpoint at $x^9=0$.

The tachyon configuration is accompanied by a gauge field configuration in the 
unbroken $O(8)$ gauge group such that
\be
\int_{S^8} \mbox{Tr}\, F^4 = 1 \,.
\ee
Note however that the gauge bundle is classified by 
$\pi_7(O(8))=\mathbb{Z} \oplus \mathbb{Z}$. There seem to be two
inequivalent charges that the endpoint of the 1-brane can carry.
We will come back to this point and its implication at the end of section 3.

\subsection{other branes}
Let's try to extend Polchinski's construction to the other D-branes in Type I string theory.
The branes which might be able to end on 9-branes are D1, D5, D7 and D8.
Interestingly there are exactly four Hopf maps,
\be
\label{Hopf}
\begin{array}{cccccc}
\mbox{0th}:  & S^0 & \rightarrow & S^1 & \rightarrow  & S^1 \\[5pt]
\mbox{1st}: & S^1 & \rightarrow & S^3  & \rightarrow & S^2\\[5pt]
\mbox{2nd}: & S^3  & \rightarrow  & S^7  & \rightarrow & S^4 \\[5pt]
\mbox{3rd}: & S^7 &  \rightarrow & S^{15}  & \rightarrow  & S^8 \,,
\end{array}
\ee
which naturally correspond to an open 8-brane, 7-brane, 5-brane and 1-brane, respectively.
The construction is a simple generalization of the 1-brane case. 
We begin with $2k$ 9-branes and $k$ anti-9-branes,
{\em i.e.} $O(2k)\times O(k)$.
The tachyon is therefore a map from $R^k$ to $R^{2k}$, and its vev
reduces to a map from $S^{k-1}$ to $S^{2k-1}$.
For $k=1,2,4$ and 8, there exists a nontrivial bundle
\be
S^{k-1}\rightarrow S^{2k-1} \rightarrow S^k \,,
\ee
which can be used to construct the asymptotic form of the tachyon
for an open $(9-k)$-brane,
\be
\label{tachyon_k}
T(\Omega) = {r+ \sum_{i=1}^{k+1}\gamma^i x^i\over\sqrt{2r(r+x^{k+1})}} T_0 \,.
\ee
However this raises a couple of puzzles. 
The topological class of this bundle takes values in $\pi_{k-1}(O(k))$,
where the $k-1$ refers to the base $S^k$ and the $O(k)$ refers to the fiber $S^{k-1}$.
In the four cases above,
\be
\label{typeI_homotopy}
\pi_0(O(1))=\mathbb{Z}_2 \quad
\pi_1(O(2))=\mathbb{Z} \quad
\pi_3(O(4))=\mathbb{Z}\oplus\mathbb{Z} \quad
\pi_7(O(8))=\mathbb{Z}\oplus\mathbb{Z} \,.
\ee
The problem is that, other than the $k=1$ case, these homotopy groups do not agree with
the corresponding D-brane charges.

In fact the same problem shows up in the construction of the infinite D$p$-branes as tachyonic solitons
in a system with $k$ 9-brane-anti-9-brane pairs. 
The resulting charges take values in precisely the groups shown in (\ref{typeI_homotopy}).
In that case we know that in order to get reliable answers we need to take a larger
number $N$ of brane-antibrane pairs, so that the homotopy groups are stable.
This follows from the correct identification of D-brane charge in K-theory \cite{Witten:1998cd}.

We propose that the same is true here.
With $N+k$ 9-branes and $N$ anti-9-branes the tachyon vev defines a map
\be
\label{tachyon_map}
S^{N-1}\rightarrow S^{N+k-1} \,.
\ee
In general $S^{N+k-1}$ cannot be obtained by fibring $S^{N-1}$
over $S^k$. This is only true for the four Hopf maps (\ref{Hopf}).
But a more general class of sphere bundles exists
\be
S^{N-1}\rightarrow M^{N+k-1}\rightarrow S^k \,,
\ee
where the total space $M^{N+k-1}$ is not in general a sphere.
These bundles are classified by $\pi_{k-1}(O(N))$, and therefore agree
precisely, for large $N$, with the D-brane charges.
The non-triviality of the bundle shows that a $p$-brane can only
end on precisely $k=9-p$ 9-branes.

We do not have an explicit expression for the asymptotic form of the
tachyon in this more general case, but we can suggest a possible way
to construct the map (\ref{tachyon_map}). 
Start with the inclusion map of the bundle $S^{N-1}\rightarrow M^{N-k-1}$,
and then construct a map $M^{N-k-1} \rightarrow S^{N-k-1}$ by mapping the
neighborhood of a point in $M^{N-k-1}$ to $S^{N-k-1}-\{0\}$, and the complement of the
neighborhood to $\{0\}$.
The tachyon is the composite map of these two maps.\footnote{We thank 
Edward Witten for suggesting this to us.}

%The first puzzle is that there are two copies of $\mathbb{Z}$ in the D1
%and D5 cases. This is easiest to see in the D5 case: since the gauge group
%is $O(4)\sim SU(2)\times SU(2)$, there are two inequivalent embedings of the $SU(2)$ instanton. 
%The second puzzle is that the charge associated with the open D7-brane is $\mathbb{Z}$
%rather than $\mathbb{Z}_2$ valued.
%The charge of the open D8-brane comes out right:
%the gauge group $O(1)=Z_2$ is discrete, and the charge associated with the end
%is given by a discrete holonomy.

\subsection{Type IIB}

In Type IIB string theory we expect to see 1-branes, 3-branes, 5-branes and
7-branes ending on 9-branes. Starting again with $2k$ 9-branes and $k$ anti-9-branes,
the gauge group is $U(2k)\times U(k)$, and the tachyon vev defines 
a map from $S^{2k-1}$ to $S^{4k-1}$.
There are now three Hopf maps
\be
S^{2k-1}\rightarrow S^{4k-1} \rightarrow S^{2k} \,,
\ee
with $k=1,2$ and 4. This gives correspondingly an open 7-brane, 5-brane and 1-brane.
The charges come out right in these cases since the groups $\pi_{2k-1}(U(k))=\mathbb{Z}$ are stable,
and we don't need to add more 9-brane-anti-9-brane pairs.
On the other hand there is no Hopf map for the 3-brane.
We are therefore forced to consider the more general class of unitary sphere bundles.
In particular the open 3-brane can be constructed as
\be
S^5 \rightarrow M^{11} \rightarrow S^6 \,.
\ee
More generally, a configuration of $N+k$ 9-branes and $N$ anti-9-branes
will give rise to an open $(9-2k)$-brane as the bundle
\be 
S^{2N-1}\rightarrow M^{2(N+k)-1} \rightarrow S^{2k} \,.
\ee
These bundles are classified by $\pi_{2k-1}(U(N))=\mathbb{Z}$, which agrees with the D-brane charges.

%We can also start with a 9-brane-anti-9-brane system with a gauge group
%$Sp(N+k)\times Sp(N)$, by considering the other orientifold projection of 
%Type IIB string theory. {\bf Puzzle: the symplectic sphere bundles}
%\be
%S^{4N-1}\rightarrow M^{4(N+k)-1}\rightarrow  S^{4k} 
%\ee
%{\bf only see the 5-brane (k=1) and the 1-brane (k=2).
%But we know there's also a 3-brane and a 4-brane,
%since $\pi_5(Sp(n))=\pi_4(Sp(n))=\mathbb{Z}_2$.}

%%%%%%%%%%%%%%%%%%%%%%%%%%
%%%%%%%%%%%%%%%%%%%%%%%%%%

%%%%%%%%%%%%%%%%%%%%%%%%%%%%%%%%%%%%%%

\section{Higgs-Tachyon model}
%\section{Breaking D-branes}

Let us now go back to the description of D-branes as topological solitons
in a system with an equal number of 9-branes and anti-9-branes, 
and consider the effect of adding extra 9-branes.
For $N$ brane-antibrane pairs the effective field theory has a gauge group 
$G_N\times G_N$, where $G_N$ is either $O(N)$, $U(N)$ or $Sp(N)$, 
depending on the theory, and a bi-fundamental tachyon field.
The lower dimensional $p$-branes are classified
by the homotopy groups of the tachyon vacuum manifold 
$\pi_{8-p}((G_N\times G_N)/G_N)=\pi_{8-p}(G_N)$.
For $U(N)$ this gives all the odd $p$ BPS D-branes of Type IIB,
and for $O(N)$ it gives the BPS 1-brane and 5-brane, and the stable
non-BPS 0-brane, 7-brane, and 8-brane. The $Sp(N)$ case corresponds to the 
other orientifold projection of Type IIB \cite{Sugimoto:1999tx}, 
in which case we get a BPS 1-brane and 5-brane, and a stable non-BPS 3-brane and 4-brane.
%However, to get the correct charges we must take the number of brane-antibrane
%pairs $N$  to be large enough so that this homotopy group is stable, {\em i.e.} 
%independent of $N$.
%The reason for this is that D-brane, or RR, charge really takes values in K-theory, 
%which is related to the stable homotopy groups. For example the D-brane charges
%in Type IIB string theory take values in 
%\be 
%K(S^{9-p}) = \pi_{8-p}(U) \,.
%\ee
%One then finds, as expected, the BPS D$p$-branes with
%$p=0,2,4,6$ and 8  in Type IIA,
%$p=1,3,5$ and 7 in Type IIB, and $p=1,5$ in Type I, as well as stable non-BPS
%D$p$-branes with $p=0,7$ and 8 in Type I.

With $k$ extra 9-branes
the gauge group is $G_{N+k}\times G_{N}$, and the subgroup which is left unbroken
by the tachyon is $G_k \times G_N$. The resulting vacuum manifold 
$G_{N+k}/G_k$ satisfies (see the appendix)
%This is the real, complex, or quaternionic Stiefel manifold $V_{N+k,N}$,
%depending on whether $G$ is $O$, $U$, or $Sp$, respectively.
%The homotopy groups of these manifolds are summarized in the appendix.
%The relevant result for us is that 
\be 
\pi_{8-p}(G_{N+k}/G_k) = 0
\ee
for $p>8-d$, where $d=k$, $2k+1$ and $4k+3$ in the $O$, $U$ and $Sp$
cases, respectively.
It therefore appears that the $p$-branes become topologically unstable
one-by-one as we increase $k$.
For each $p$-brane there is a critical number of excess 9-branes $k$
at which it loses its topological charge.
For example the soliton describing the D1-brane in Type I string theory
becomes topologically unstable when $k=8$.\footnote{Of course the tadpole
condition fixes $k=32$ in Type I so none of the solitons are topologically stable
\cite{Bergman:2000tm}, but the statement is still correct.}
This is precisely the number of extra 9-branes that were needed for the construction
of the semi-infinite 1-brane in the previous section. 
In fact the smallest value of $k$ for any $p$-brane soliton to become topologically unstable
is the same as that needed for the construction of the semi-infinite
$p$-brane. This suggests that the $p$-brane becomes topologically unstable
as a soliton because it can break open on the 9-branes.
As we will now show, this is related to the existence of a $(p-1)$-dimensional 
soliton.

\subsection{a simple field theory example}

Consider a four-dimensional $SU(2)$ gauge theory with two scalar fields,
a triplet and a doublet. Assume that both have a nonzero vev, and that
the triplet vev is larger than the doublet vev.
The gauge symmetry is broken in two stages:
\be
SU(2) \rightarrow U(1) \rightarrow 0 \,.
\ee
The second stage by itself is the Abelian Higgs model, 
which admits a Nielsen-Olesen string soliton with charge in $\pi_1(U(1))=\mathbb{Z}$.
This string confines the magnetic flux of the broken $U(1)$.
The first stage is the 'tHooft-Polyakov model, which gives rise 
to a magnetic monopole with charge in $\pi_2(SU(2)/U(1))=\mathbb{Z}$.
The full theory has no topologically stable solitons since 
$\pi_1(SU(2))=\pi_2(SU(2))=0$. The magnetic flux of the monopole
from stage 1 gets confined into a semi-infinite string in stage 2.
Alternatively, the string from stage 2 can break by the creation
of a monopole-anti-monopole pair in stage 1. 
Either way we see that the loss of topological charge
is explained by the configuration of a string ending on a monopole.

\subsection{Type II}

The simple field theory example above illustrates our strategy for
the brane-anti-brane system. We first separate the extra branes from
the brane-antibrane pairs by turning on a vev for the appropriate adjoint scalars,
and then let the tachyon condense.\footnote{Since we are using 9-branes we can't
really separate them, but we can compactify them and turn on Wilson lines.
This is T-dual to separating lower-dimensional branes.}
We will call the first stage the "Higgs stage", and the second the "tachyon stage".
In the Type IIB 9-brane-anti-9-brane system this gives
\be
U(N+k) \times U(N)  \rightarrow U(k) \times U(N) \times U(N)
\rightarrow U(k) \times U(N)\,.
\ee
The vacuum manifold of the full theory is given by
\be
{\cal M}_k = {U(N+k)\over U(k)} \,,
\ee
which has trivial homotopy groups $\pi_{8-p}({\cal M}_k)=0$ for $p\geq 8-2k$.
The vacuum manifolds of the Higgs and tachyon stage are given by
\be
{\cal M}_k^{H} &=& {U(N+k)\over U(k) \times U(N)} \\
{\cal M}_k^{T} &=& U(N) \,.
\ee
The homotopy groups of the three manifolds above have different physical 
meanings. We already saw that $\pi_{8-p}({\cal M}_k)$ classifies $p$-dimensional
solitons in the full theory. The group $\pi_{8-p}({\cal M}_k^{T}$) classifies
$p$-dimensional solitons in the tachyon stage, which is the brane-antibrane annihilation
stage. These are precisely the $p$-branes. Finally, the group 
$\pi_{9-p}({\cal M}_k^{H})$ classifies $(p-1)$-dimensional solitons in the Higgs stage.
In the simple example above we saw that the existence of a 
pointlike soliton in the first stage is what allowed the string-like soliton of the second stage to break.
We shall therefore refer to the solitons in the Higgs stage as {\em end-solitons}.
In general a $p$-brane will be able to break if there exists at least one matching
$(p-1)$-dimensional end-soliton in the Higgs stage. 

The homotopy groups of the three manifolds form an exact sequence given by
\be
\label{exact_sequence}
\cdots \rightarrow \pi_{9-p}({\cal M}_k) \stackrel{j*}{\rightarrow} \pi_{9-p}({\cal M}_k^{H}) 
\stackrel{\partial *}{\rightarrow}
\pi_{8-p}({\cal M}_k^{T}) 
\stackrel{i*}{\rightarrow} \pi_{8-p}({\cal M}_k) \rightarrow \cdots \,.
\ee
This implies that when $\pi_{8-p}({\cal M}_k)=0$ the map
\be
\label{homotopy_map}
\pi_{9-p}({\cal M}_k^{H}) \stackrel{\partial *}{\rightarrow}  \pi_{8-p}({\cal M}_k^{T}) 
\ee 
is onto (surjective), because 
$\mbox{Im}\, \partial * = \mbox{Ker}\, i* = \pi_{8-p}({\cal M}_k^{T})$.
In other words there is at least one end-soliton for each $p$-brane
with $p\geq 8-2k$. Therefore 
the $p$-brane can break precisely
when it is no longer stable as a soliton in the full theory, which is what we 
wanted to demonstrate.

Since there are only odd-dimensional stable $p$-branes in Type IIB
to begin with, the smallest $p$-brane that can break has $p=9-2k$.
For $p\geq 9-2k$ the exact sequence actually shows that the map 
(\ref{homotopy_map}) is an isomorphism, since in this case
both $\pi_{8-p}({\cal M}_k)=\pi_{9-p}({\cal M}_k)=0$.
This means there is only one possible end-soliton for each $p$-brane.
This will not be the case in Type I.

The $(p-1)$-dimensional end-soliton
has the correct charge in the unbroken gauge group $U(k)$
to satisfy the condition of section 2,
\be
\int_{S^{9-p}} \mbox{Tr}\, F^{(9-p)/2} \in \pi_{8-p}(U(k)) = \mathbb{Z} \;\;
\mbox{for}\;\; p\geq 9-2k \,.
\ee
As a consistency check we note that the gauge charge is identical to
the topological charge of the end-soliton in $\pi_{9-p}({\cal M}_k^{H})$
thanks to identity (\ref{Grassman2}).
Figure~1 shows a 7-brane breaking on a 9-brane (or a 1-brane
on a 3-brane).

\begin{figure}
\centerline{\includegraphics[width=3.5in]{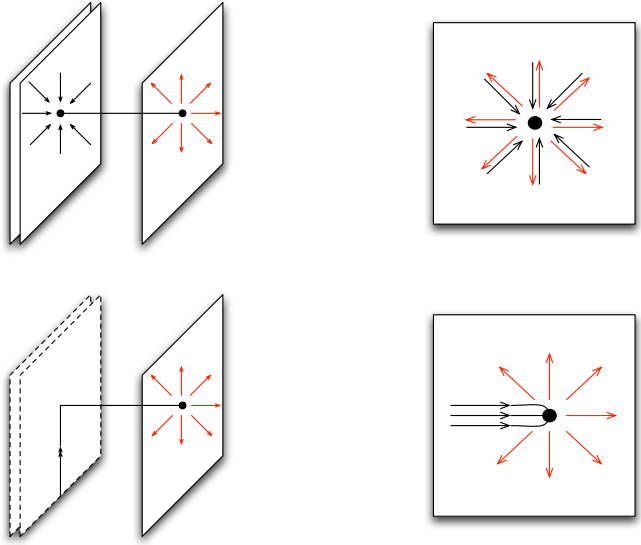}}
\caption{A $p$-brane breaking on $(p+2)$-branes from the spacetime and worldvolume
points of view. On top an 'tHooft-Polyakov monopole forms when we separate the brane-antibrane
pair from the extra brane. On the bottom the magnetic flux on the brane-antibrane is confined
to a semi-infinite $p$-brane after the tachyon condenses.}
\end{figure}

\subsection{Type I}

Applying the same strategy in Type I string theory, 
we break the 9-brane-anti-9-brane gauge symmetry in two stages 
by first separating the extra 9-branes (turning on the appropriate Wilson line),
and then letting the tachyon condense,
\be
O(N+k) \times O(N)  \rightarrow O(k) \times O(N) \times O(N)
\rightarrow O(k) \times O(N)\,.
\ee
The three vacuum manifolds in this case are given by
\be
{\cal M}_k &=& {O(N+k)\over O(k)} \\
{\cal M}_k^{H} &=& {O(N+k)\over O(k)\times O(N)} \\
{\cal M}_k^{T} &=& O(N) \,.
\ee
The solitons of the full theory are classified by $\pi_{8-p}({\cal M}_k)$,
which in this case vanishes for $p\geq 9-k$. 
We therefore expect the 8-brane to first break when $k=1$,
the 7-brane when $k=2$, the 5-brane when $k=4$, and the 1-brane
when $k=8$.
This is confirmed, exactly as in the Type IIB case,
by looking at the exact homotopy sequence (\ref{exact_sequence}).
When $\pi_{8-p}({\cal M}_k)=0$, the map 
$\pi_{9-p}({\cal M}_k^{H}) \rightarrow \pi_{8-p}({\cal M}_k^{T})$
is onto, and therefore each $p$-brane with $p\geq 9-k$ has at least
one end-soliton associated to it.

For $p\geq 10-k$ the map is an isomorphism, so there is exactly one
end-soliton for each $p$-brane.
However now the smallest $p$-brane that can break has $p=9-k$,
for which the map is onto, but not one-to-one.
The end-solitons in these cases are not in one-to-one correspondence with
the $p$-branes. 

This can have an interesting effect. 
Consider a finite open $p$-brane with two ends.
If the group of end-solitons is isomorphic
to the group of $p$-branes, the two ends carry opposite
charges and the configuration is unstable.
However if the two groups are not isomorphic,
it should be possible to have an open $p$-brane
with ends whose charges do not sum up to zero.
Let's look at each of the four $p$-branes in turn.

\medskip

\noindent\underline{\em 1-brane} :
For the 1-brane with $k=8$ the particle-like end-solitons are classified by
\be
\label{one_brane}
\pi_8({\cal M}_8^{H}) = \mathbb{Z} \oplus \mathbb{Z} \,,
\ee
whereas the 1-brane itself is in $\pi_7({\cal M}_8^{T})=\mathbb{Z}$.
There are two inequivalent types of end-solitons.
Therefore an open 1-brane with one end on one type of
soliton, and the other end on the other type of (anti-) soliton
should be stable.
Indeed, the full theory contains a stable particle-like soliton in this case,
since
\be
\pi_8({\cal M}_8) = \mathbb{Z} \,.
\ee

\medskip

\noindent\underline{\em 5-brane}:
The 5-brane with $k=4$ can likewise break in two inequivalent ways,
since $\pi_4({\cal M}_4^{H}) = \mathbb{Z} \oplus \mathbb{Z}$.
This leads to a stable 4-dimensional soliton, as can be seen from
$\pi_4({\cal M}_4) = \mathbb{Z}$.
%This is also consistent with the fact that the gauge bundle is an $SU(2)$
%instanton, which can be embedded in two distinct ways into $O(4)\sim SU(2)\times SU(2)$.
%The stable particle of (\ref{particle}) is really a D1-brane with one end charged under
%the first $SU(2)$ and the other end charged under the second $SU(2)$.\footnote{A similar
%thing happens in the D1-D9 case with $k=8$. Since 
%$\pi_7(O(8))=\mathbb{Z} \oplus \mathbb{Z}$ the D1-brane has two topologically
%distinct endpoints in the D9-brane. However we cannot implement the two-stage
%symmetry breaking in this case, since we cannot separate the D9-branes.}

\medskip

\noindent\underline{\em 7-brane}:
For the 7-brane with $k=2$ the possible end-solitons are classified by
\be
\pi_2({\cal M}_2^{H}) = \mathbb{Z} \,,
\ee
whereas the 7-brane is in $\pi_1({\cal M}_2^{T})=\mathbb{Z}_2$.
In this case there is a single type of end-soliton, but since the 7-brane
does not have an orientation (it is its own antibrane) the two endpoints 
can be charged equally or oppositely. In the former case the object should 
be stable. This is confirmed by the existence of a stable 6-dimensional soliton
in
\be
\pi_2({\cal M}_2) = \mathbb{Z} \,.
\ee
Furthermore, from the part of the exact sequence
\be
\begin{array}{ccccc}
\pi_2({\cal M}_2) & \rightarrow & \pi_2({\cal M}_2^{H}) & \rightarrow &
\pi_1({\cal M}_2^{T}) \\
\mathbb{Z} & \rightarrow & \mathbb{Z} & \rightarrow & \mathbb{Z}_2
\end{array}
\ee
we see that the basic soliton carries two units of end-soliton charge.

\medskip

\noindent\underline{\em 8-brane}:
In the case of the 8-brane with $k=1$ the end-solitons carry a
$\mathbb{Z}_2$ charge, and are therefore in one-to-one correspondence
with the 8-brane. Correspondingly,
there is no stable 7-dimensional soliton in the full theory.

\medskip

Finally, as in the Type II case, the end-solitons also carry the required
charges in the unbroken $O(k)$
\be
\int_{S^{9-p}} \mbox{Tr}\, F^{(9-p)/2} \in \pi_{8-p}(O(k))\,,
\ee
which agree with their topological charge in $\pi_{9-p}({\cal M}_k^{H})$
thanks to (\ref{Grassman2}).\footnote{We are assuming here that there is a proper 
generalization of the RR charge conservation argument which led to (\ref{end_charge})
for the case of the Type I 7-brane and 8-brane, whose RR fields exist only as torsion
in K-theory. It would be interesting to formulate this in a precise manner.}

%%%%%%%%%%%%%%%%%%%%%%%%%%%%%%%%%%%%%%%%
%%%%%%%%%%%%%%%%%%%%%%%%%%%%%%%%%%%%%%%%

\section{Discussion}

We have given two descriptions of $p$-branes ending on 9-branes.
The two descriptions agree on the conditions for branes to end,
but they are valid in different regimes.
In the tachyon model all the 9-branes and anti-9-branes coincide
(no Wilson line), so the broken $p$-brane is inside the branes it ends on.
In the Higgs-Tachyon model we first break the symmetry by separating the extra
9-branes from the brane-anti-brane pairs, and then let the tachyon condense.
The $p$-brane forms at the location of the pairs, and breaks by connecting to
a $p$-brane which stretches between the pairs and the extra 9-branes (Fig.~1).
The former is the $p$-dimensional soliton formed in the tachyon
stage, and the latter is the $(p-1)$-dimensional end-soliton formed in the Higgs stage.

An interesting question is whether we can connect the two descriptions
by gradually reducing the Higgs vev (or really the Wilson line in the 9-brane case), 
{\em i.e.} the separation between
the brane-anti-brane pairs and the extra 9-branes.
As we reduce the Higgs vev the core of the end-soliton grows, until it fills
all of space, at which point the full gauge group $G_{N+k}\times G_N$ would be unbroken 
everywhere. This clearly does not fit with the first description, in which the gauge group
is broken to $G_k\times G_N$ almost everywhere.
Furthermore, the semi-infinite part of the $p$-brane carries different fluxes
in the two descriptions. In the Higgs-Tachyon model the $p$-brane carries
the magnetic flux of the gauge group which is broken in the tachyon stage,
whereas in the tachyon model it carries the unbroken flux.
Apparently a phase transition must occur between the two descriptions.

%In this note we described the breaking of a D-brane in two different regimes.
%The description using topological solitons in the two stage symmetry breaking
%is relevant for large Higgs vev, for which (in a geometrical picture)
%the broken brane touches the higher
% dimensional brane at an angle close to $pi/2$, and stays somewhat outside 
%of it. The second description using 
%Polchinskies construction is relevant for a brane which is broken with-in the 
%higher dimesnional brane.

%We have seen that both 
%constructions agree on the circumstances when branes can break. In both cases 
%the breaking branes are clasified by $\pi_{k}(SO(N))$, and the end points by
%$\pi_{k-1}(SO(k))$.

%One may 
%imagine that we can start with the first situation and try to lower the 
%separation between the broken brane and the higher dimensional brane, and 
%arrive at the second construction in a smooth way, but this is not so. 

%In the
% first construction the broken brane is made out of confining flux of the 
%broken group, and the end point is a generalised t' Hooft-Polyakov momopole, 
%in which at the core the broken symmetry is restored. the size of the monopole
%is inversly proportional to the Higgs vev and so trying to lower the Higgs 
%vev to zero in a smooth way will result in an unbroken gauge group almost 
%every where. This clearly does not fit the construction by Polchinski, 
%thus there must be some phase transition on the way.

%\medskip

We found three examples of topologically stable open $p$-branes
in 9-branes with orthogonal groups: a 1-brane in eight 9-branes,
a 5-brane in four 9-branes, and a 7-brane in two 9-branes.
In these cases there are more types of end-solitons than $p$-branes,
which allows the sum of the two charges at the ends to be non-zero.
Of course tadpole cancellation requires an excess of exactly 32 9-branes,
so these states aren't really stable in Type I string theory.
One could break the $SO(32)$ symmetry with additional Wilson lines
and obtain states which would be stable in some energy range.
Alternatively one could imagine (via T-duality for example) lower-dimensional
orientifold planes, in which these states would be stable.

%%%%%%%%%%%%%%%%%%%%%%%%%%%%%%%%
%%%%%%%%%%%%%%%%%%%%%%%%%%%%%%%%

\section*{Acknowledgments}
We wish to thank Joe Polchinski for patiently explaining his results,
and for the initial correspondence which led to this project.
We would also like to thank Edward Witten for useful discussions.
This work is supported in part by the
Israel Science Foundation under grant no.~568/05.

%%%%%%%%%%%%%%%%%%%%%%%%%%%%%%%%
%%%%%%%%%%%%%%%%%%%%%%%%%%%%%%%%

\appendix
\section{Various homotopy groups}

The homotopy groups used in this paper can be found in \cite{Steenrod}.
The stable homotopy groups of the simple Lie groups are given by
\be
\pi_i(U(n)) = \left\{
\begin{array}{ll}
\mathbb{Z} & i \;\;\mbox{odd} \\
0 & i \;\;\mbox{even}
\end{array}
\right.
\ee
for $i<2n$,
\be
\label{orthogonal_homotopy}
\pi_i(O(n)) = \left\{
\begin{array}{ll}
\mathbb{Z} & i=3,7 \;\;\mbox{mod}\; 8\\
\mathbb{Z}_2 & i=0,1 \;\;\mbox{mod}\; 8\\
0 & i=2,4,5,6 \;\;\mbox{mod}\; 8
\end{array}
\right.
\ee
for $i<n-1$, and
\be
\pi_i(Sp(n)) = \left\{
\begin{array}{ll}
\mathbb{Z} & i=3,7 \;\;\mbox{mod}\; 8\\
\mathbb{Z}_2 & i=4,5 \;\;\mbox{mod}\; 8\\
0 & i=0,1,2,6 \;\;\mbox{mod}\; 8
\end{array}
\right.
\ee
for $i<4n+2$.
Some relevant unstable homotopy groups are
\be
\pi_1(O(2))=\pi_5(O(6))=\mathbb{Z} \; , \; \pi_3(O(4))=\pi_7(O(8))=\mathbb{Z} \oplus \mathbb{Z} \,.
\ee
The vacuum manifolds in the Higgs-Tachyon model included the Stiefel manifolds
\be
V_{n,m} = {G_n\over G_{n-m}} \,,
\ee
which are real, complex and quaternionic for the cases 
when $G_n$ is $O(n)$, $U(n)$ and $Sp(n)$, respectively, and
the Grassman manifolds
\be
M_{n,m} = {G_n\over G_{n-m}\times G_m} \,.
\ee
The homotopy groups of the Stiefel manifolds $V_{n,m}$ satisfy
\be
\label{Stiefel_zero}
\pi_i(V_{n,m})=0 \;\; \mbox{for} \;\; i <  \left\{
\begin{array}{ll}
n-m & \mathbb{R} \\
2(n-m)+1 & \mathbb{C} \\
4(n-m)+3 & \mathbb{H}
\end{array}
\right. \,.
\ee
At the upper limit of the above ranges the homotopy groups are given by
\be
\label{Stiefel_real}
\pi_{n-m}(V^{\mathbb{R}}_{n,m}) = \left\{
\begin{array}{ll}
\mathbb{Z} & n-m \;\; \mbox{even} \\
\mathbb{Z}_2 & n-m \;\; \mbox{odd}
\end{array}
\right. \,,
\ee
and 
\be
\label{Stiefel_complex}
\pi_{2(n-m)+1}(V^{\mathbb{C}}_{n,m}) = \pi_{4(n-m)+3}(V^{\mathbb{H}}_{n,m}) = \mathbb{Z}\,.
\ee
There is an exact sequence relating the homotopy groups of $G_m$, $V_{n,m}$
and $M_{n,m}$:
\be
\label{sequence}
\cdots \rightarrow \pi_i(V_{n,m}) \rightarrow \pi_i(M_{n,m}) \rightarrow
\pi_{i-1}(G_{m}) \rightarrow \pi_{i-1}(V_{n,m}) \rightarrow \cdots \,.
\ee
Using (\ref{Stiefel_zero}) it then follows that
\be
\label{Grassman1}
\pi_i(M_{n,m}) = \pi_{i-1}(G_{m}) \;\; \mbox{for} \;\; i <  \left\{
\begin{array}{ll}
n-m & \mathbb{R} \\
2(n-m)+1 & \mathbb{C} \\
4(n-m)+3 & \mathbb{H}
\end{array}
\right. \,.
\ee
For $n-m<m$ a more useful identity is obtained using the equivalence
$M_{n,m}=M_{n,n-m}$:
\be
\label{Grassman2}
\pi_i(M_{n,m}) = \pi_{i-1}(G_{n-m}) \;\; \mbox{for} \;\; i < \left\{
\begin{array}{ll}
m & \mathbb{R} \\
2m+1 & \mathbb{C} \\
4m+3 & \mathbb{H}
\end{array}
\right. \,.
\ee

%%%%%%%%%%%%%%%%%%%%%%%%%%%%%
%%%%%%%%%%%%%%%%%%%%%%%%%%%%%

%%%%%%%%%%%%%%%%%%%%%%%%%%%
%%%%%%%%%%%%%%%%%%%%%%%%%%%


\begin{thebibliography}{99}

%\cite{Strominger:1995ac}
\bibitem{Strominger:1995ac}
  A.~Strominger,
  %``Open p-branes,''
  Phys.\ Lett.\ B {\bf 383}, 44 (1996)
  [arXiv:hep-th/9512059].
  %%CITATION = HEP-TH 9512059;%%


 %\cite{Copeland:2003bj}
\bibitem{Copeland:2003bj}
  E.~J.~Copeland, R.~C.~Myers and J.~Polchinski,
  %``Cosmic F- and D-strings,''
  JHEP {\bf 0406}, 013 (2004)
  [arXiv:hep-th/0312067].
  %%CITATION = HEP-TH 0312067;%%


%\cite{Polchinski:2005bg}
\bibitem{Polchinski:2005bg}
  J.~Polchinski,
  %``Open heterotic strings,''
  arXiv:hep-th/0510033.
  %%CITATION = HEP-TH 0510033;%%

%\cite{Hashimoto:2001rj}
\bibitem{Hashimoto:2001rj}
  K.~Hashimoto and S.~Hirano,
  %``Branes ending on branes in a tachyon model,''
  JHEP {\bf 0104}, 003 (2001)
  [arXiv:hep-th/0102173].
  %%CITATION = HEP-TH 0102173;%%

%\cite{Sen:1998tt}
\bibitem{Sen:1998tt}
  A.~Sen,
  %``SO(32) spinors of type I and other solitons on brane-antibrane pair,''
  JHEP {\bf 9809} (1998) 023
  [arXiv:hep-th/9808141].
  %%CITATION = HEP-TH 9808141;%%
  
  
%\cite{Witten:1998cd}
\bibitem{Witten:1998cd}
  E.~Witten,
  %``D-branes and K-theory,''
  JHEP {\bf 9812}, 019 (1998)
  [arXiv:hep-th/9810188].
  %%CITATION = HEP-TH 9810188;%%


%\cite{Hashimoto:2005qh}
\bibitem{Hashimoto:2005qh}
  K.~Hashimoto and S.~Terashima,
  %``ADHM is tachyon condensation,''
  JHEP {\bf 0602}, 018 (2006)
  [arXiv:hep-th/0511297].
  %%CITATION = HEP-TH 0511297;%%
  
%\cite{Ishida:2006tj}
\bibitem{Ishida:2006tj}
  A.~Ishida, S.~Uehara and T.~Yada,
  %``Tachyon condensation in unbalanced D anti-D system,''
  arXiv:hep-th/0601050.
  %%CITATION = HEP-TH 0601050;%%

%\cite{Sugimoto:1999tx}
\bibitem{Sugimoto:1999tx}
  S.~Sugimoto,
  %``Anomaly cancellations in type I D9-D9-bar system and the USp(32)  string
  %theory,''
  Prog.\ Theor.\ Phys.\  {\bf 102}, 685 (1999)
  [arXiv:hep-th/9905159].
  %%CITATION = HEP-TH 9905159;%%
  

%\cite{Bergman:2000tm}
\bibitem{Bergman:2000tm}
  O.~Bergman,
  %``Tachyon condensation in unstable type I D-brane systems,''
  JHEP {\bf 0011}, 015 (2000)
  [arXiv:hep-th/0009252].
  %%CITATION = HEP-TH 0009252;%%
  
  
\bibitem{Steenrod}
N.~Steenrod,
"The topology of fibre bundles,"
Princeton University Press 1951.
  
 %%%%%%%%%%%%%%%%%%%%%%%%%%%%%
 
  
  
  
  
  

\end{thebibliography}
\end{document}